\documentstyle[a4,11pt]{article}

\newcommand{\nit}{\noindent}

\newcommand{\np}{\newpage}
\newcommand{\dsp}{\displaystyle}
\newcommand{\vs}[1]{\vspace{#1 ex}}
\newcommand{\hs}[1]{\hspace{#1 em}}
\newcommand{\bfr}{\begin{flushright}}
\newcommand{\efr}{\end{flushright}}
\newcommand{\bc}{\begin{center}}
\newcommand{\ec}{\end{center}}
\newcommand{\ben}{\begin{enumerate}}
\newcommand{\een}{\end{enumerate}}

\newcommand{\be}{\begin{equation}}
\newcommand{\ee}{\end{equation}}
\newcommand{\ba}{\begin{array}}
\newcommand{\ea}{\end{array}}
\newcommand{\ct}{\cite}
\newcommand{\bit}{\bibitem}
\newcommand{\dd}[2]{\frac{\partial{#1}}{\partial{#2}}}
\newcommand{\ag}{\alpha}
\newcommand{\bg}{\beta}

\newcommand{\del}{\delta}

\newcommand{\ve}{\varepsilon}

\newcommand{\thg}{\theta}

\newcommand{\rg}{\rho}

\newcommand{\vf}{\varphi}
\newcommand{\og}{\omega}
\newcommand{\Gam}{\Gamma}
\newcommand{\Del}{\Delta}
\newcommand{\Fg}{\Phi}

\newcommand{\tT}{\tilde{T}}

\newcommand{\lh}{\left(}
\newcommand{\rh}{\right)}

\newcommand{\tg}{\mbox{\,tg\,}}
\newcommand{\ctg}{\mbox{\,ctg\,}}

\newcommand{\der}{\partial}

\begin{document}

\pagestyle{empty}
\begin{flushright}
NIKHEF 01-007
\end{flushright}

\begin{center}
{\Large{ \bf{Particles, fluids and vortices} }}\\
\vs{5}

{\large J.W.\ van Holten} \\
\vs{2}

{\large{NIKHEF, Amsterdam NL}} \\
\vs{2}
{\tt t32@nikhef.nl}
\vs{2}

July 17, 2001\\
\vs{10}

{\small{ \bf{Abstract} }} \\
\end{center}

\nit
{\footnotesize{Classical particle mechanics on curved spaces is related
to the flow of ideal fluids, by a dual interpretation of the Hamilton-Jacobi
equation. As in second quantization, the procedure relates the description
of a system with a finite number of degrees of freedom to one with infinitely
many degrees of freedom. In some two-dimensional fluid mechanics models a
duality transformation between the velocity potential and the stream function
can be performed relating sources and sinks in one model to vortices in the
other. The particle mechanics counterpart of the dual theory is reconstructed.
In the quantum theory the strength of sources and sinks, as well as vorticity
are quantized; for the duality between theories to be preserved these
quantization conditions must be related. }} \vfill

\np
\pagestyle{plain}
\pagenumbering{arabic}

\section{Particles \label{s1}}

The free motion of a classical particle with unit mass, moving in a smooth
space with metric $g_{ij}(x)$ is described by the Lagrangean
\be
L = \frac{1}{2}\, g_{ij}(x) \dot{x}^i \dot{x}^j,
\label{1.1}
\ee
where as usual the overdot represents a time-derivative. The Euler-Lagrange
equations imply that the particle moves on a geodesic:
\be
\frac{D^2 x^i}{Dt^2} =
 \ddot{x}^i + \Gam_{jk}^{\;\;\;\;i} \dot{x}^j \dot{x}^k = 0.
\label{1.2}
\ee
The canonical formulation of this theory is constructed in terms of the momenta
\be
p_i = \frac{\der L}{\der \dot{x}^i} = g_{ij} \dot{x}^j,
\label{1.3}
\ee
and the hamiltonian
\be
H = \frac{1}{2}\, g^{ij} p_i p_j.
\label{1.4}
\ee
The time-development of any scalar function $F(x,p)$ of the phase-space
co-ordin\-ates is then determined by the Poisson brackets
\be
\frac{dF}{dt} = \left\{ F, H \right\} = \frac{\der F}{\der x^i}
 \frac{\der H}{\der p_i} - \frac{\der F}{\der p_i} \frac{\der H}{\der x^i}.
\label{1.5}
\ee
In particular the Hamilton equations themselves read
\be
\dot{x}^i = \frac{\der H}{\der p_i}, \hs{2}
\dot{p}_i = - \frac{\der H}{\der x^i}.
\label{1.6}
\ee
A third formulation of the classical theory is provided by Hamilton's
principal function\footnote{The terminology follows ref.\ct{goldstein}.}
$S(x,t)$, which is the solution of the partial differential equation
\be
\frac{\der S}{\der t}\, = - H(x, p= \nabla S).
\label{1.7}
\ee
For the case at hand this Hamilton-Jacobi equation takes the form
\be
\frac{\der S}{\der t}\, = - \frac{1}{2}\, g^{ij}\, \nabla_i S\, \nabla_j S.
\label{1.8}
\ee
Particular solutions $S$ are provided by the action for classical paths
$x^i(\tau)$ obeying the Euler-Lagrange equation (\ref{1.2}), starting at
time $\tau = 0$ at an initial point $x^i(0)$, and reaching the point
$x^i(t)=x^i$ at time $\tau = t$:
\be
S(x,t) = \left. \int_0^t d\tau L(x,\dot{x})\, \right|_{x^i(\tau)}.
\label{1.9}
\ee
An example of the class of theories of this type is that of a particle
moving on the surface of the unit sphere, $S^2$. A convenient co-ordinate
system is provided by the polar angles $(\thg, \vf)$, in terms of which
\be
L(\thg, \vf) = \frac{1}{2}\, \lh \dot{\thg}^2 + \sin^2 \thg\, \dot{\vf}^2
 \rh,
\label{1.10}
\ee
for a particle of unit mass. The corresponding hamiltonian is
\be
H = \frac{1}{2}\, \lh p_{\thg}^2 + \frac{p_{\vf}^2}{\sin^2 \thg} \rh =
 \frac{{\bf J}^2}{2},
\label{1.11}
\ee
with the momenta and velocities related by
\be
p_{\thg} = \dot{\thg}, \hs{2} p_{\vf} = \sin^2 \thg\, \dot{\vf}.
\label{1.12}
\ee
The second equality (\ref{1.11}) relates the hamiltonian to the Casimir invariant
of angular momentum, the components of which are constants of motion given by
\be
J_x = - \sin \vf\, p_{\thg} - \cos \vf \ctg \thg\, p_{\vf}, \hs{1.5}
J_y = \cos \vf\, p_{\thg} - \sin \vf \ctg \thg\, p_{\vf}, \hs{1.5}
J_z = p_{\vf}.
\label{1.13}
\ee
The geodesics on the sphere are the great circles; they can be parametrized by
\be
\cos \thg(\tau) = \sin \ag \sin \og (\tau - \tau_*), \hs{2}
 \tg(\vf(\tau) - \vf_*) = \cos \ag\, \tg \og (\tau - \tau_*),
\label{1.14}
\ee
where $\ag$ is a constant, and $\tau_*$ and $\vf_*$ are the time and
longitude at which the orbit crosses the equator: $\thg_* = \pi/2$.
On these orbits the angular frequency is related to the total angular
momentum by
\be
\og^2 =  2H = {\bf J}^2,
\label{1.14.0}
\ee
Observe that, for an orbit reaching the point with co-ordinates $(\thg, \vf)$
at time $\tau_* + t$, the following relations hold:
\be
\ba{l}
\cos \og = \sin \thg\, \cos (\vf - \vf_*), \hs{2}
\sin \og t = \sqrt{ 1 - \sin^2 \thg\, \cos^2 (\vf - \vf_*) }, \\
 \\
\sin \ag = \dsp{ \frac{\cos \thg}{\sqrt{ 1 - \sin^2 \thg \cos^2 (\vf - \vf_*) }}.}
\ea
\label{1.14.1}
\ee
The last equation implicitly describes the orbit $\thg(\vf)$, defining a great
circle which cuts the equator at $\thg = \thg_* = \pi/2$ and $\vf = \vf_*$,
at an angle $\ag$ defined by the direction of the angular momentum:
\be
\frac{J_z}{\sqrt{ {\bf J}^2 }} = \cos \ag, \hs{2} \frac{J_{\perp}}{\sqrt{
 {\bf J}^2 }} = \sin \ag, \hs{1} J_{\perp} = \sqrt{J_x^2 + J_y^2}.
\label{1.14.3}
\ee
The Hamilton-Jacobi equation for this system reads
\be
\frac{\der S}{\der t} = - \frac{1}{2}\, \left[ \lh \frac{\der S}{\der \thg}
 \rh^2 + \frac{1}{\sin^2 \thg} \lh \frac{\der S}{\der \vf}\rh^2 \right].
\label{1.15}
\ee
The solution corresponding to the orbit (\ref{1.14}) is
\be
S(\thg,\vf,t) = \frac{1}{2t}\,
 \arccos^2 \left[ \sin \thg\, \cos (\vf - \vf_*) \right],
\label{1.16}
\ee
which satisfies the equations
\be
\ba{lll}
\dsp{ \frac{\der S}{\der \thg} }& = & \dsp{ p_{\thg} = -
 \frac{\og \cos \thg\, \cos(\vf - \vf_*)}{\sqrt{ 1 - \sin^2 \thg
 \cos^2 (\vf - \vf_*) }} , }\\
 & & \\
\dsp{ \frac{\der S}{\der \vf} }& = & \dsp{ p_{\vf} =
 \frac{\og \sin \thg\, \sin(\vf - \vf_*)}{\sqrt{ 1 - \sin^2 \thg
 \cos^2 (\vf - \vf_*) }}, }\\
 & & \\
\dsp{ \frac{\der S}{\der t} }& = & \dsp{ - H = - \frac{\og^2}{2}.}
\ea
\label{1.17}
\ee
In this approach, the expressions on the right-hand side are obtained by
{\em defining} $\og$ via the last expression, in agreement with (\ref{1.14.1}).
The same principle of energy conservation/time-translation invariance implies
that $S$ does not depend on $\tau_*$.

\section{Fluids \label{s2}}

The Hamilton-Jacobi equation (\ref{1.8}) can itself be obtained in a
straightforward way from a variational principle: introduce a Lagrange
multiplier field $\rg(x)$ and construct the
action functional
\be
A(\rg,S) = \int dt \int d^n x\, \sqrt{g}\, \rg \lh \dd{S}{t} + \frac{1}{2}\, g^{ij}\,
 \nabla_i S\, \nabla_j S \rh.
\label{2.1}
\ee
The square root of the (time-independent) background metric has been included
to make the integration measure invariant under reparametrizations. Of course,
we could absorb it in the definition of Lagrange multiplier field, but then
$\rg$ would transform as a density rather than as scalar.

The Hamilton-Jacobi equation follows by requiring the action to be
stationary w.r.t.\ variations of $\rg$:
\be
\frac{1}{\sqrt{g}}\, \frac{\del A}{\del \rg} = \dd{S}{t} +
 \frac{1}{2}\, g^{ij}\, \nabla_i S\, \nabla_j S = 0.
\label{2.2}
\ee
On the other hand, the stationarity of $A(\rg,S)$ w.r.t.\ $S$ gives
\be
- \frac{1}{\sqrt{g}}\, \frac{\del A}{\del S} = \dd{\rg}{t} +
  \nabla_i \lh g^{ij} \rg \nabla_j S \rh = 0.
\label{2.3}
\ee
This equation can be interpreted as the covariant equation of continuity for
a fluid\footnote{For background, see e.g.\ ref.\ct{ll}.} with density $\rg$
and local velocity
\be
v_i = \nabla_i S \hs{1} \Rightarrow \hs{1}
 \dd{\rg}{t} + \nabla_i \lh \rg v^i \rh = 0.
\label{2.4}
\ee
In this interpretation the gradient of the Hamilton-Jacobi equation gives the
covariant Euler equation
\be
\dd{v_i}{t} + v^j \nabla_j v_i = 0, \hs{2}
\nabla_j v_i = \dd{v_i}{x^j} - \Gam_{ji}^{\;\;\;\;k} v_k.
\label{2.5}
\ee
Eq.(\ref{2.4}) states that the fluid flow is of the potential type.
Indeed, in the absence of torsion the Riemann-Christoffel connection
$\Gam_{ij}^{\;\;\;\;k}$ is symmetric and  the local vorticity vanishes:
\be
\nabla_i v_j - \nabla_j v_i = 0.
\label{2.7}
\ee
For the fluid flow to be incompressible, the velocity field must be divergence
free:
\be
\nabla \cdot v = \Del S = 0,
\label{2.8}
\ee
where $\Del = g^{ij}\, \nabla_i \nabla_j$ is the covariant laplacean on
scalar functions over the space. It follows that the number of incompressible
modes of flow on the manifold equals the number of zero-modes of the scalar
laplacean. For example, for flow on the sphere $S^2$ (or any other compact
Riemann surface) there is only one incompressible mode, the trivial one with
$v^i = 0$ everywhere.

For a given geometry $g_{ij}(x)$, the solution of the Hamilton-Jacobi equation
(\ref{1.8}), (\ref{2.2}) provides a special solution of the Euler equation
(\ref{2.5}); for a conservative system: $\der S/\der t = -H =$ constant, it
implies $\der v_i / \der t = 0$ and $v^j \nabla_j v_i = 0$. Accordingly, this
solution describes geodesic flow starting from the point $(\thg_*, \vf_*)$.

To turn this into a complete solution of the fluid-dynamical equations
(\ref{2.4}), (\ref{2.5}) it remains to solve for the density $\rg$. The
equation of continuity takes the form
\be
\dd{\rg}{t} + \nabla_i (\rg \nabla^i S) = 0.
\label{2.10}
\ee
It follows that a stationary flow, with $\rg$ not explicitly depending on
time $t$, is possible if
\be
\nabla \cdot (\rg \nabla S) = 0.
\label{2.11}
\ee
In addition to the trivial solution $\rg = \rg_0 =$ constant, $v = \nabla
S/m = 0$, it is possible to find non-trivial solutions of equation
(\ref{2.11}) for spatially varying density $\rg$. As an example, we consider
flow in a 2-dimensional space; in this case one can introduce a generalized
stream function $T(x,t)$, dual to the fluid momentum, and write
\be
\rg \nabla^i S = \frac{1}{\sqrt{g}}\, \ve^{ij} \nabla_j T.
\label{2.11.1}
\ee
Then for theories of the type (\ref{1.8}):
\be
\rg = \frac{\ve^{ij} \nabla_i S \nabla_j T}{\sqrt{g} (\nabla S)^2}\,
 =\, \frac{\ve^{ij} \nabla_i S \nabla_j T}{2 H \sqrt{g}}.
\label{2.11.2}
\ee
With $H$ constant, the factor $2H$ in the denominator can be absorbed
into the definition of $\tT = T/2H$, and hence the density is given by
\be
\rg = \frac{1}{\sqrt{g}}\, \ve^{ij} \nabla_i S \nabla_j \tT
    = \frac{1}{\sqrt{g}}\, \ve^{ij} v_i \nabla_j \tT ,
\label{2.11.3}
\ee
for the pseudo-scalar function $T$ the gradient of which is dual to
$\rg \nabla S$. Note also, that eq.(\ref{2.11.1}) implies $\nabla S \cdot
\nabla T = v \cdot \nabla T = 0$.
\vs{1}

\nit
As an illustration, we again consider the unit sphere $S^2$. The velocity
field is given by the momenta (\ref{1.17}) per unit mass:
\be
v_{\thg} = - \frac{\og \cos \thg\, \cos(\vf - \vf_*)}{
 \sqrt{ 1 - \sin^2 \thg \cos^2 (\vf - \vf_*) }}, \hs{2}
v_{\vf} = \frac{\og \sin \thg\, \sin(\vf - \vf_*)}{
 \sqrt{ 1 - \sin^2 \thg \cos^2 (\vf - \vf_*) }}.
\label{2.11.4}
\ee
Taking into account that on the sphere the non-vanishing components
of the connection are
\be
\Gam_{\thg \vf}^{\;\;\;\;\vf} = \frac{\cos \thg}{\sin \thg}, \hs{2}
\Gam_{\vf\vf}^{\;\;\;\;\thg} = - \sin \thg \cos \thg,
\label{2.13}
\ee
a straightforward calculation shows that indeed
\be
v_j v^j = \og^2, \hs{2} v^j \nabla_j v_i = 0, \hs{2} \dd{v_i}{t} = 0.
\label{2.14}
\ee
The first two equations actually imply $v^j (\nabla_i v_j - \nabla_j v_i) = 0$,
in agreement with the absence of local circulation. From these results
it follows, that the flowlines are geodesics (great circles) given by
eq.(\ref{1.14.1}), and stationary.

For the gradient of the stream function $T$ to be orthogonal to the velocity
field (\ref{2.11.4}), it must satisfy the linear differential equation
\be
v \cdot \nabla T = 0 \hs{1} \Leftrightarrow \hs{1} \tg (\vf - \vf_*)\,
 \nabla_{\vf} T = \sin \thg\, \cos \thg\, \nabla_{\thg} T.
\label{2.14.1}
\ee
The general solution can be obtained by separation of variables, and is a
function of the single variable: $T(\thg,\vf) = f\lh y \rh$, with $y = \tg \thg\,
\sin (\vf - \vf_*) = \ctg \ag$. For such a scalar field
\be
\nabla_{\thg} T = \frac{\sin (\vf - \vf_*)}{\cos^2 \thg}\, \left.
 f^{\prime}(y)\right|_{y = \tiny{\ctg} \ag}, \hs{2} \nabla_{\vf} T = \tg \thg\,
 \cos(\vf - \vf_*)\, \left. f^{\prime}(y)\right|_{y = \tiny{\ctg} \ag}.
\label{2.16}
\ee
The corresponding density $\rg$ is then
\be
\rg(\thg,\vf) = \frac{\bar{\rg}(\ag)}{\cos \thg}\, =\, - \frac{1}{\og \sin \ag
 \cos \thg}\, \left. f^{\prime}(y)\right|_{y = \tiny{\ctg} \ag}.
\label{2.17}
\ee
The simplest, most regular solution is obtained for $\bar{\rg}(\ag) = \rg_*
\sin \ag$:
\be
\rg(\thg,\vf) = \frac{\rg_* \sin \ag}{\cos \thg} =
 \frac{\rg_*}{\sqrt{ 1 - \sin^2 \thg \cos^2 (\vf -\vf_*)}}.
\label{2.18}
\ee
This solution corresponds to
\be
T(\thg,\vf) = \og \rg_*\, \ag(\thg,\vf) \hs{1} \Leftrightarrow \hs{1}
 f(y) = \og \rg_* \mbox{\,arcctg\,} y.
\label{2.18.1}
\ee
Observe, that in this case $T$, like $\ag$, is an angular variable; indeed,
$\ag$ increases by $2 \pi n$ on any loop winding around the point $(\thg =
\pi/2; \vf = \vf_*)$ $n$ times.

The solution (\ref{2.18}) possesses singular points at $\thg = \pi/2$,
$\vf = \vf_* + n \pi$, corresponding to a source for $n = 0$, and a sink
for $n = 1$. This can be established from the expression for $\nabla \cdot v$:
\be
\nabla \cdot v = \frac{\og \sin \thg \cos (\vf - \vf_*)}{
\sqrt{1 - \sin^2 \thg \cos^2 (\vf - \vf_*)}},
\label{2.19}
\ee
which becomes $(+\infty, -\infty)$ at the singular points. However, a more
elegant way to establish the result, is to make use of the stream function
(\ref{2.18.1}) and consider the flux integral
\be
\Fg(\Gam) = \oint_{\Gam} dl\, \rg v_n,
\label{2.20}
\ee
representing the total flow of material across the closed curve $\Gam$ per
unit of time. Consider a contour $\Gam$ winding once around the singularity
at $(\thg = \pi/2; \vf = \vf_*)$; on such a curve $\ag$ increases from $0$
to $2\pi$. Then
\be
\Fg(\Gam) = \oint_{\Gam} \sqrt{g} \ve_{ij} \rg v^i dx^j =
 \oint_{\Gam} \nabla_i T dx^i = 2\pi \og \rg_*.
\label{2.21}
\ee
This represents the total flow of matter from the hemisphere centered on the
source at $(\thg = \pi/2; \vf = \vf_*)$ to the hemisphere centered on its
antipodal point, the sink at $(\thg = \pi/2; \vf = \vf_* + \pi)$.

\section{Vortices \label{s3}}

The dual relationship between the velocity potential $S$ and the stream
function $T$ suggests to study the dynamics of a fluid for which $T$ is the
velocity potential:
\be
v_i = \frac{1}{\rg_*}\, \nabla_i T.
\label{3.1}
\ee
The constant $\rg_*$ has been included for dimensional reasons.
Like before, this velocity field is stationary: $\der v_i/\der t = 0$, but
it is not geodesic. Indeed, the velocity field describes motion under the
influence of an external potential; specifically:
\be
v \cdot \nabla v_i = \frac{1}{2}\, \nabla_i\, v^2 =
 \frac{1}{2\rg_*^2}\, \nabla_i (\nabla T)^2 =
 \frac{1}{2 \rg_*^2}\, \nabla_i (\rg \nabla S)^2.
\label{3.2}
\ee
Here $\rg(x)$ and $S(x)$ denote the previously defined functions mapping
the manifold to the real numbers ---e.g.\ (\ref{1.16}) and (\ref{2.18})
for fluid motion on a sphere--- irrespective of their physical
interpretation. Now again, as $(\nabla S)^2 = 2 H = \og^2 =$ constant,
it follows that
\be
v \cdot \nabla v_i = \frac{\og^2}{2\rg_*^2}\, \nabla_i\, \rg^2 \equiv -
 \nabla_i h.
\label{3.3}
\ee
Combining eqs.(\ref{3.2}) and (\ref{3.3}):
\be
\frac{1}{2}\, v^2 = - (h - h_0) = \frac{\og^2 \rg^2}{2 \rg_*^2},
\label{3.4}
\ee
where $h$ represents the external potential. Because of the potential nature
of the flow, eq.\ (\ref{3.1}), the local vorticity again vanishes: $\nabla_i
v_j - \nabla_j v_i = 0$,  but as eq.(\ref{2.21}) shows, this is not necessarily
true globally. Indeed, in singular points of the original geodesic fluid flow
(with sources/sinks), the dual flow generally has vortices/anti-vortices.

Continuing our example from the previous sections, we can illustrate these
results in terms of flow on the unit sphere, for which $T/\rg_* =
\og \ag$ and $v_i = \og \nabla_i \ag$:
\be
v_{\thg} = - \frac{\og \sin (\vf - \vf_*)}{1 - \sin^2 \thg \cos^2 (\vf - \vf_*)},
 \hs{2}
v_{\vf} = - \frac{\og \sin \thg \cos \thg \cos(\vf - \vf_*)}{1 - \sin^2 \thg
 \cos^2 (\vf - \vf_*)}.
\label{3.5}
\ee
It follows, as expected, that
\be
v^2 = \og^2 (\nabla \ag)^2 = \frac{\og^2 \rg^2}{\rg_*^2}\, =
 \frac{\og^2}{1 - \sin^2 \thg \cos^2 (\vf - \vf_*)}.
\label{3.6}
\ee
A further remarkable property, is that the dual flow is divergence free:
\be
\nabla \cdot v = 0 \hs{1} \Leftrightarrow \hs{1} \Del \ag = 0,
\label{3.7}
\ee
where-ever $v$ is well-defined; obviously, the result can only be
true because of the two singular points $(\thg = \pi/2; \vf = \vf_*)$
and $(\thg = \pi/2; \vf = \vf_* + \pi)$, where $v_i$ and its divergence
are not well-defined, i.e.\ topologically the velocity field is defined
on a cylinder, rather than a sphere. These two points are centers of
vorticity, as follows directly from eq.(\ref{2.21}), which in the present
context can be rewritten as
\be
\oint_{\Gam} v_i dx^i = 2 \pi \og,
\label{3.8}
\ee
for any closed curve $\Gam$ winding once around the singular point
$(\pi/2,\vf_*)$; as this curve also winds once around the other singular
point in the opposite direction, they clearly define a pair of vortices of
equal but opposite magnitude.

As the flow is divergence free, it follows that in this case there can be
non-trivial incompressible and stationary flow modes: for constant
density $\rg_1$ one has
\be
\dd{\rg_1}{t} = 0, \hs{2} \nabla \rg_1 = 0 \hs{1} \Rightarrow \hs{1}
 \nabla \cdot (\rg_1 v) = 0,
\label{3.9}
\ee
and the equation of continuity is satisfied.

The nature of the flow lines defined by eq.(\ref{3.5}) is clear: they are
parallel circles of equidistant points around the centers of vorticity.
On these circles the velocity is constant in magnitude, implying by
(\ref{3.6}) that $\sin \thg \cos(\vf - \vf_*) \equiv \cos \bg =$ constant.
For example, for $\vf_* = 0$ we get $x = \cos \bg =$ constant; the flow
line then is the circle where this plane of constant $x$ cuts the unit
sphere. On these flow lines
\be
v_{\thg} = - \og_1 \sin (\vf - \vf_*), \hs{2}
v_{\vf} = - \og_1 \cos \bg \cos \thg,
\label{3.10}
\ee
with
\be
\og_1 = \frac{v^2}{\og}\, = \frac{\og}{1 - \sin^2 \thg \cos^2 (\vf - \vf_*)}\,
 = \frac{\og}{\sin^2 \bg}.
\label{3.11}
\ee

\section{The dual particle model \label{s4}}

Having clarified the nature of the (incompressible) flow described by
the dual velocity potential $T/\rg_*$, we now reconstruct the corresponding
particle-mechanics model for which $T/\rg_*$ is Hamilton's principal
function. From eqs.(\ref{3.4}), (\ref{3.6}) we observe that the hamiltonian
is of the form $H_1 = K + h$, with for the specific case at hand a
kinetic-energy term:
\be
K = \frac{1}{2}\, g^{ij} p_i p_j\, \rightarrow\,
 \frac{1}{2}\, \lh p_{\thg}^2 + \frac{p_{\vf}^2}{\sin^2 \thg} \rh,
\label{4.2}
\ee
and the potential (normalized for later convenience such that $2H = \og\og_1$):
\be
h(\thg, \vf) = h_0 - \frac{\og^2 \rg^2}{2 \rg_*^2}\, \rightarrow\,
 \frac{\og\og_1}{2} \lh 1 - \frac{\og/\og_1}{1 - \sin^2 \thg \cos^2
 (\vf - \vf_*)} \rh.
\label{4.3}
\ee
The corresponding lagrangean $L_1 = K - h$ produces the Euler-Lagrange equations
\be
\ba{lll}
\dot{p}_{\thg} & = & \dsp{\ddot{\thg}\, =\, \sin \thg \cos \thg\, \dot{\vf}^2
 + \frac{\og^2 \sin \thg \cos \thg\, \cos^2 (\vf - \vf_*)}{(1 - \sin^2 \thg
 \cos^2 (\vf - \vf_*) )^2}, }\\
 & & \\
\dot{p}_{\vf} & = & \dsp{ \frac{d}{dt} \lh \sin^2 \thg\, \dot{\vf} \rh\, =\,
 - \frac{\og^2 \sin^2 \thg \sin (\vf - \vf_*) \cos (\vf - \vf_*)}{(1 -
 \sin^2 \thg \cos^2 (\vf - \vf_*) )^2}. }
\ea
\label{4.5}
\ee
These equations have solutions
\be
\cos \thg = \sin \bg\, \sin \og_1 t, \hs{2}
\tg (\vf - \vf_*) = \tg \bg\, \cos \og_1 t,
\label{4.6}
\ee
with $\bg$ a constant, implying the relation
\be
\sin \thg\, \cos (\vf - \vf_*) = \cos \bg.
\label{4.7}
\ee
Solving for the velocity (and taking into account the unit mass)
\be
p_{\thg} = v^{\thg} = - \og_1 \sin (\vf - \vf_*), \hs{2}
p_{\vf} = \sin^2 \thg\, v^{\vf} = - \og_1 \cos \bg\, \cos \thg,
\label{4.8}
\ee
in agreement with (\ref{3.10}). From these results we can compute Hamilton's
principal function
\be
S_1(\thg,\vf,t) = \int_0^t d\tau\, L_1[\thg(\tau),\vf(\tau)] = \frac{1}{2t}\,
 \mbox{arcctg}^2 \lh \tg \thg \sin (\vf - \vf_*) \rh.
\label{4.9}
\ee
This function indeed satisfies the Hamilton-Jacobi equations
\be
\dd{S_1}{\thg} = p_{\thg}, \hs{2} \dd{S_1}{\vf} = p_{\vf},
\label{4.11}
\ee
with $(p_{\thg}, p_{\vf})$ as given by eq.(\ref{4.8}), and
\be
\dd{S_1}{t} =\, -\frac{\og\og_1}{2} = -\frac{1}{2}\, \left[ \lh \dd{S_1}{\thg}
 \rh^2 + \frac{1}{\sin^2 \thg} \lh \dd{S_1}{\vf} \rh^2 \right] - h(\thg,\vf).
\label{4.10}
\ee
Using the relation $\ctg \ag = \tg \thg \sin (\vf - \vf_*) = \ctg \og_1 t$,
the equations (\ref{4.11}) can be recast in the form
\be
p_i = \og \nabla_i \ag = \frac{1}{\rg_*}\, \nabla_i T.
\label{4.12}
\ee
Hence $T/\rg_*$ can indeed be identified with Hamilton's principal function
of this system.

Repeating the arguments of sect.\ \ref{s2}, the action (\ref{2.1}) for the
Hamilton-Jacobi theory is now generalized to:
\be
A(\rg,S_1;h) = \int dt \int d^n x\, \sqrt{g}\, \rg \lh \dd{S_1}{t} + \frac{1}{2}\,
 g^{ij} \nabla_i S_1 \nabla_j S_1 + h \rh.
\label{4.13}
\ee
Reinterpretation of $S_1$ as a velocity potential for fluid flow: $v = \nabla
S_1$, leads back directly to the inhomogeneous Euler equation
\be
\dd{v_i}{t} + v \cdot \nabla v_i = - \nabla_i h,
\label{4.14}
\ee
which for stationary flow becomes eq.(\ref{3.3}). Variation of this
action w.r.t.\ $S_1$ gives the equation of continuity for $\rg$, as before;
note that in this action $h$ plays the role of an external source for the
density $\rg$.

\section{Quantum theory \label{s5}}

The quantum theory of a particle on a curved manifold is well-established.
For the wave function to be well-defined and single-valued, the momenta must
satisfy the Bohr-Sommerfeld quantization conditions
\be
\oint_{\Gam} p_i dx^i = 2 \pi n \hbar,
\label{5.1}
\ee
for any closed classical orbit $\Gam$. For the free particle of unit mass
on the unit sphere the left-hand side is
\be
\int_0^T v^2 d\tau = \og^2 T = 2 \pi \og,
\label{5.2}
\ee
where $T = 2\pi/\og$ is the period of the orbit. Hence the quantization
rule amounts to quantization of the rotation frequency (the angular momentum):
$\og = n \hbar$.

For the dual model, the same quantity takes the value
\be
\oint_{\Gam} v_i dx^i = \int_0^{T_1} v^2 d\tau = \frac{\og^2 T_1}{\sin^2 \bg}\,
 = \og\, \og_1 T_1 = 2 \pi \og,
\label{5.4}
\ee
and again $\og = n \hbar$. As the quantization conditions in the two dual
models are the same, the duality can be preserved in the quantum theory.

If this is to be true also in the fluid interpretation, the quantization
conditions must be respected at that level as well. Now the first quantization
condition for the integral (\ref{5.2}) is interpreted in the fluid dynamical
context as a quantization of the fluid momentum, cf.\ eq.(\ref{2.11.4}). The
second quantization condition (\ref{5.4}) has a twofold interpretation: first,
according to eqs.(\ref{2.20}), (\ref{2.21}) it quantizes the strength of the
fluid sources and sinks in the model of free geodesic flow; the agreement
between the two quantization conditions is then obvious: in order for the
strength of the source/sink to satisfy a quantization condition, the amount
of fluid transfered from one to the other must be quantized as well.

In the context of the dual model however, the condition imposes the
quantization of vorticity in the quantum fluid \ct{feynman}. In the more
general context of quantum models of fluids in geodesic flow on a compact
two-dimensional surface and their duals described by the stream functions,
this observation shows that duality at the quantum level requires the
quantization of sources in one model to be directly related to
the quantization of vorticity in the dual one. This situation
closely parallels the relation between the quantization of monopole charge
\ct{dirac} and the quantization of the magnetic flux of fluxlines
\ct{abrikosov} in three dimensions.

\end{document}